\documentstyle[12pt]{article}

\newcommand{\dteta}{\partial_\theta}
\newcommand{\be}{\begin{equation}}
\newcommand{\ee}{\end{equation}}

\newcommand{\beqn}{\begin{eqnarray}}
\newcommand{\eeqn}{\end{eqnarray}}

\begin{document}

\title{On $q$-Deformed Supersymmetric Classical Mechanical Models}
\author{$L. P.~ Colatto^{a,b}$\thanks{%
email: colatto@ictp.trieste.it}~~ $and$~~ $J.L.~ Matheus-Valle^{b,c}$\thanks{%
e-mail: zeluiz@cbpfsu1.cat.cbpf.br} \\
\\
$^a$International Centre for Theoretical Physics, \\
Strada Costiera, 11, P. O. Box 586, 34100, Trieste, Italy; \\
\\
$^b$CBPF/CNPq, Rua Dr. Xavier Sigaud, 150 \\
22290-180 Rio de Janeiro, RJ, Brazil \\
and \\
$^c$Departamento de F\'\i sica, ICE \\
Universidade Federal de Juiz de Fora \\
36036-330 Juiz de Fora, MG, Brazil}
\date{}
\maketitle
\thispagestyle{empty}
\abstract{\ Based on the idea of quantum groups and
paragrassmann  variables, we present a generalization of supersymmetric
classical mechanics with a deformation parameter $q= \exp{\frac{2 \pi i}{k}}$
dealing  with the $k =3$ case. The coordinates of the $q$-superspace are a
commuting parameter $t$ and a paragrassmann variable $\theta$, where $%
\theta^3 =  0$. The generator and covariant derivative are obtained, as well
as the  action for some possible superfields. }

\begin{center}
MIRAMARE - TRIESTE \\October 1995
\end{center}

\newpage

\setcounter{page}{1}
%
%

\section{Introduction}

In the last few years, Quasi Triangular Hopf Algebras or Quantum Groups \cite
{Skl,Dri,Wor} have attracted a lot of attention from  physicists. One of the
most interesting features is that such structures can be related to
underlying symmetries on spaces where the coordinates are non-commutative
\cite{Wess}.

It has been shown that the creation and annihilation operators of the $q$%
-deformed harmonic oscillator \cite{Mac}
\begin{equation}
a\,a^{\dagger} -q a^{\dagger} \, a = q^{-N},  \label{acomadagar}
\end{equation}
possess a classical limit where these operators can be understood as
coordinates  obeying \cite{Bau}
\begin{equation}
\theta^k = 0,
\end{equation}
where $k$ is an integer, and the $q$-factor of the deformation is a prime
root of unity, $q^k=1$. In general, the properties of these coordinates are
generalizations of the associated with  Grassmann variables. Promoting these
coordinates to functions of a (non-deformed) parameter $t$, it was shown
that it is possible to write down an action for such fields that, when added
to the action of a commuting field, has a symmetry resembling supersymmetry
\cite{Nos1}, and it has also been how to functional integrate on a heterotic
quantum field theory \cite{noB300}. The aim of this letter is to show how a
way to understand the transformations on such fields, and the action
invariances, as resulting from a superspace formulation of a classical
mechanical model where its coordinates are the paragrassmann variables (a $q$%
-superspace), and non-commuting fields.

In the next section we briefly review paragrassmann variables and also how
we construct coordinates and actions from them. Section 3 is devoted to the
construction of the $q$-superspace, transformations between its coordinates,
and the induced transformations on the $q$-superfields defined on it.
Invariant quadratic actions are constructed in Section 4, in particular for
a free particle and the harmonic oscillator. We leave some final comments to
the last section.

%
%

\section{Paragrassmann Variables and their Relation to Quer\-mionic
Coordinates}

We start this section by introducing a paragrassmann variable $\theta$  and
its derivative, $\frac{\partial}{\partial \theta} \equiv  \partial_{\theta}$
obeying \cite{Fil}
\begin{equation}
\theta^k = 0 ,\mbox{~~~~~~~} {\partial_{\theta}}^k = 0,  \label{tetaak}
\end{equation}
for a positive integer $k$.

If we demand that the action of $\partial_\theta $ on $\theta^n$ is
proportional to $\theta^{n-1}$, it turns out that it becomes necessary to
deform the Leibnitz rule to be
\begin{equation}
\partial_\theta (a\,b) = (\partial_\theta a) b + g(a) (\partial_\theta b),
\label{dab}
\end{equation}
where $a,b$ are arbitrary polynomials in $\theta$, and $g(a)$ is an
automorphism of the algebra, satisfying
\begin{eqnarray}
g(\alpha \,a + \beta \,b) &=& \alpha \, g(a) + \beta \, g(b),  \nonumber \\
g(a\, b) &=& g(a) \, g(b),
\end{eqnarray}
where $\alpha, \beta$ are c-numbers.

Choosing $a = \theta$ in (\ref{dab}), we see that $\partial_\theta$ and $%
\theta$ must obey a $q$-deformed commutation (quommutation) relation
\begin{equation}
[  \dteta, \theta]_q \equiv \dteta \theta - q \theta \dteta = 1 ,
\label{dcomteta}
\end{equation}
implying for $\theta$ the automorphism
\begin{equation}
g(\theta) = q \, \theta.
\end{equation}

This derivative, however, is not unique. Indeed, we could change the power $1
$ in eq.~(\ref{dcomteta}) by any other integer, thus for each value of $k$
one can define $k-1$ different derivatives. For the specific case $k=3$, one
may also define another derivative $\delta _\theta $ \cite{Ahn} that
quommutes with $\theta $ as
\begin{equation}
\lbrack \delta _\theta ,\theta ]_{q^2} \equiv \delta _\theta \theta -q^2\theta
\delta _\theta =1,
\end{equation}
and its Leibnitz rule differs from eq.(\ref{dab}) by changing $g(a)$ to $%
g(g(a))$. These two derivatives have the following $q$-commutation relation

$$\partial _\theta \delta _\theta -q\delta _\theta \partial _\theta =1$$

As in the Grassmannian case, it is not possible to define the integral over $%
\theta $ as the inverse of the derivative. However, if we impose translation
invariance and homogenity for the integral, it must be of the form
\begin{equation}
\int d\theta \,\theta ^n\;\alpha \;\delta _{n,k-1}.  \label{int}
\end{equation}

It is interesting to notice that, for $k=2$, $q=-1$, eq.(\ref{acomadagar})
becomes the usual anticommutator, consistent with eqs.(\ref{tetaak}) and (%
\ref{dcomteta}), which are the conditions for Grassmann variables.  Taking $%
k \rightarrow \infty$, eq.(\ref{acomadagar}) becomes the usual commutator.
The meaning of this limit in eq.(\ref{tetaak}) is  that, if we Taylor expand
a function of these variables, it will become a series (obviously, if $%
\theta^k = 0$, $k$ finite, a Taylor expansion will be a polynomial of degree
$(k-1)$), being
\begin{equation}
f= a_0 + \theta a_1 + \theta^2 a_2 + ... + \theta^{k-1} a_{k-1}
\end{equation}
where we can promote the functions of a (commuting) parameter $t$.

 Let us recall that in the
Grassmannian case we have two different coordinates: one that behaves like $%
\theta$ (a fermionic coordinate), and another that behaves like $\theta^0$,
a bosonic (commuting) coordinate. In the paragrassmann case, we will have $k$
different types of coordinates, each one  corresponding to a power of $\theta
$, and again $\theta^0$ being a commuting one. We  call $\psi^{(i)}(t)$
the $q$-fermionic generalization of the coordinates or, simply, the
quermionic coordinates and its label $(i)$ indicates the sector to which it
belongs.

In a some  recent work \cite{Nos1}, it was emphasized that two quermions of
different sectors  obey the quommutation relation
\begin{equation}
\psi^{(i)} \, \psi^{(j)} = q_{(i,j)} \psi^{(j)} \, \psi^{(i)},  \label{qpsi}
\end{equation}
where the parameters $q_{(i,j)}$ are simply powers of $q$, $q^k = 1$.

The particular case $k=3$ was taken, and an action which extends the
supersymmetric point particle through the use of these generalized fields
was constructed. This generalized particle was described by the coordinates $%
( x(t), \psi^{(1)}(t), \psi^{(2)}(t)) $, in the same way as a supersymmetric
point particle is described by the coordinates $(x(t), \psi(t)) $. The
showed action involving the quermions was given by
\begin{equation}
S = \int dt (\frac{1}{2} \dot{x}^2 - q C^{(s)^2} {\dot \psi}^{(2)}
\psi^{(1)} ) ,  \label{act}
\end{equation}
with the mass equal to one. The second term in (\ref{act}) resembles the
classical fermionic equation of motion. The cocycle--type factor $C^{(s)^2}$
was required because the product of two objects of different sectors, $%
A^{(r)} B^{(s)}$, must behave like an object of the sector $(r+s)\, {\rm mod}%
\, 3$. In that work, it was emphasized  the necessity of the factor
Cq-superfield, paying the price of writing  a suitable ``algebra" of this
factors. For instance, the cocycle--type  factor $C^{(s)}$ that could be
seen as a sector--coun\-ter, had a relation
\begin{equation}
C^{(s)} A^{(i)} = q^i A^{(i)} C^{(s)},
\end{equation}
and adding the choice $[\psi^{(1)},\psi^{(2)}]_q=0 $, that take all the
fields as real, the second term in the action eq.~(\ref{act}) was left real
and a zeroth sector representative.

Another interesting feature to recall was the transformation (the
variations of a field, from now on, will be written as $\Delta$ to one not
to be confused with the derivative $\delta$),
\begin{eqnarray}
\Delta x & = & q C^{(s)} \epsilon^{(1)} \psi^{(2)},  \nonumber \\
\Delta \psi^{(1)} & = & q^2 C^{(s)^2} \epsilon^{(1)} \dot x ,  \nonumber \\
\Delta \psi^{(2)} & = & \pm q \epsilon^{(1)} \psi^{(1)} ,  \label{deltacomp1}
\end{eqnarray}
on the action (\ref{act}) reaching
\begin{equation}
\Delta S = \pm \int dt \frac{d }{dt} (\epsilon^{(1)} \psi^{(1)^2}),
\label{deltaact}
\end{equation}
where $[\epsilon^{(1)},\psi^{(1)}]_q = [\psi^{(2)},\epsilon^{(1)}]_q=0$ was
used. Such transformation is similar to a supersymmetric one: the parameter $%
\epsilon^{(1)}$ is a non-commuting one, the action transforms as a total
derivative, and one  of the fields, $\psi^{(1)}$, transforms as a total
derivative, which can be taken as indicating that $\psi^{(1)}$ is the
highest term in a $\theta$--expansion of some superfield. One could also
write transformations among the fields with a parameter belonging to the
sector-two. However, it can be shown that this transformation is not a
symmetry of the action (\ref{act}) \cite{Nos1}.



\section{The $q$-Superspace and $q$-Superfields}

We begin to construct a $q$-superspace formulation that will recover the
structure concerning the quermionic coordinates presented in the last
section. As previously stated, we will consider in detail only the $k=3$
case, that represents  the nilpotency and produces an interesting expression
$1+q+q^2=0$. It  emphasizes that the $q$ and $q^2$ cases have not crucial
difference. Some of the ideas discussed here and in the next section had
been discussed also in refs. \cite{Durand,Debergh} and more recently in
\cite{Mac2}\footnote{This nice work appeared when we was submitting this
paper.}.

The $q$-superspace coordinates are $(t;\theta)$, where $t$ is a c-number to
be identified with time and $\theta$ is a paragrassmannian  variable obeying
$\theta^3 = 0$, and both are taken as real parameters.

Let us now introduce transformations between these coordinates that are
translations on the $q$-superspace. We  write them as
\begin{eqnarray}
\theta^{\prime}&=& \theta + \varepsilon,  \nonumber \\
t^{\prime}&=& t + q^C \theta^2 \varepsilon ,  \label{vardetetaet}
\end{eqnarray}
where $\varepsilon$ is an infinitesimal constant in the same sector as $%
\theta$ and $C$ can assume the values $1,2,3$.  Clearly, the exponent $3$
will give us a trivial factor restricting then our set of possible
choices. The translation in $q$-superspace fixes the mass dimensions of $%
\theta$ and $\varepsilon$ to be $-\frac{1}{3}$. Although the translation
term in $t$ does not commute with the infinitesimal parameter $\varepsilon$,
it still belongs to the same sector as $t$. (Remember that we met this issue
when we wrote down the action for the quermionic components, eq.(\ref{act})
and we  introduced the cocycle-like factor $C^{(s)}$ to correct the
statistics). We will say that two terms are homogeneous if botthe same sector.
Defining the quommutator to be
\begin{equation}
[A,B]_q \equiv A\,B - q\,B\,A,  \label{quomutador}
\end{equation}
we choose
\begin{equation}
[\theta, \varepsilon]_{q^{2C}} = 0.  \label{tetacomeps}
\end{equation}
It is after determining these quommutation relations that we set the $q$
factors in (\ref{vardetetaet}) to preserve the reality condition for the
coordinates.  We could choose $q^C$ instead of $q^{2C}$ in (\ref{tetacomeps}%
) (i.e., take ${[}\varepsilon, \theta ]_{q^{2C}} = 0$). With this choice  we
necessarily have to change $q \leftrightarrow q^2$ in eq. (\ref{vardetetaet}%
).

After introducing the $q$-superspace $(t,\theta)$, our next step is to write
down a function of these variables. As in the supersymmetric case, let us
expand this function in a Taylor series on $\theta$, this expansion is a
polynomial of degree 2 (for the generic case $\theta^k=0$, the polynomial
goes up to the order $(k-1)$),
\begin{equation}
X(t;\theta) = x(t) + q^{B_2} \; \theta \;\psi^{(2)}(t) + q^{2B_1} \;
\theta^2 \; \psi^{(1)}(t) .  \label{fi}
\end{equation}
The coordinate $x(t)$ is a commuting function, the $\psi^{(i)}(t)$ are
the $q$-supersymmetric partners of the coordinate $x(t)$,
and their dimensions are ${[}\psi^{(j)}] = -\frac{j}{3}$. We take their
quommutators to be
\begin{eqnarray}
[ {\psi}^{(1)}, \psi^{(2)}{]}_{q^A} &=& 0 ,  \nonumber \\
{[} \varepsilon, \psi^{(j)}{]}_{q^{D_j}} &=& 0,  \nonumber \\
{[} \theta, \psi^{(j)}{]}_{q^{B_j}}&=& 0 , \label{alg}
\end{eqnarray}
where the last expression guarantee that $X$ is real and the others
complete a deformed algebra.

The infinitesimal coordinate transformations (\ref{vardetetaet}) induce a
variation on the $q$-superfield $X(t,\theta)$ of the form
\begin{equation}
X(t^{\prime},\theta^{\prime}) - X(t,\theta) = \Delta \, X = \varepsilon \; Q
\, X.  \label{deltafi}
\end{equation}
We can get the realization of the  q-supersymmetric generator
transformation, $Q$, by Taylor expanding the l.r.s. of this equation.
Choosing the factors to keep the reality condition we have
\begin{equation}
X(\theta^{\prime},t^{\prime}) - X(\theta,t) = \varepsilon \frac{\partial X}{%
\partial \theta} + q^{2C} \varepsilon \theta^2 \frac{\partial X}{\partial t},
\end{equation}
With this expansion, and using eq.(\ref{vardetetaet}), $Q$ becomes
\begin{equation}
Q = q^{2C} \theta^2 \frac{\partial}{\partial t} + \frac{\partial}{\partial
\theta}.
\end{equation}
We notice that the generator is in the $\theta^2$ sector, and its canonical
dimension is $[Q] = \frac{1}{3} $.  A straightforward calculation shows that
$Q^3 = -\partial_t$. This means that the $q$-supersymmetric transformations
are the cubic roots of time translations.

Explicitly computing the r.h.s. of (\ref{deltafi}), we obtain the $X$
variation as
\begin{equation}
\Delta X = q^{B_2} \varepsilon \psi^{(2)} - q^{2 B_1 + 2 +C} \theta
\varepsilon \psi^{(1)} + q^{2 C} \theta^2 \varepsilon \dot{x}.
\label{DeltaX}
\end{equation}
Bearing in mind the reality condition we find from $\Delta X$ and $X$
itself some relations among the $q$ exponents. Finally we reach
\begin{eqnarray}
2 C = 2 B_2 = B_1,  \nonumber \\
D_2 = D_1 +1.
\end{eqnarray}
The above relations do not fix completely the quommutators among the
variables (see (\ref{alg})) we are considering, since we still have at  our
disposal three free coefficients. We may choose the variables $2 C= D_2 = 1$
and $A=2$, thus  fixing all the other ones (remember that the $k=3$
nilpotency we have only two relevant  choices for the exponents). With such
a choice, the $q$-superspace  translation becomes
\begin{eqnarray}
\theta ^{\prime}= \theta + \varepsilon,  \nonumber \\
t^{\prime}= t + q^2 \theta^2 \varepsilon,
\end{eqnarray}
while the $q$-su\begin{equation}
X(t) = x + q^2 \theta \psi^{(2)} + q^2 \theta^2 \psi^{(1)}.
\end{equation}
The $q$-SUSY generator
\begin{equation}
Q = q \theta^2 \partial_t + \partial_{\theta},
\end{equation}
yields the transformation
\begin{equation}
\Delta X = q \varepsilon \theta^2 \dot x + q^2 \varepsilon \psi^{(2)} - q
\varepsilon \theta \psi^{(1)},
\end{equation}
or in components
\begin{eqnarray}
\Delta x = q^2 \varepsilon \psi^{(2)},  \nonumber \\
\Delta \psi^{(1)} = \varepsilon \dot x,  \nonumber \\
\Delta \psi^{(2)} = q \varepsilon \psi^{(1)}.  \label{deltacomp}
\end{eqnarray}
Moreover, they have the quommutators
\begin{eqnarray}
\varepsilon \theta = q^2 \theta \varepsilon,  \nonumber \\
\theta {\psi^{(j)}} = q^j {\psi^{(j)}} \theta,  \nonumber \\
\varepsilon {\psi^{(j)}} = q^j {\psi^{(j)}} \varepsilon.
\label{quomutadores}
\end{eqnarray}
which let on the same structure as the one present in Section 2. This
structure allows us to take the quommutation relation between the two
quermionic coordinates, which read
\begin{equation}
\psi^{(1)} \psi^{(2)} = q^2 \psi^{(2)} \psi^{(1)}.
\end{equation}

Having written down the $q$-superspace transformations and the variations on
the $q$-superfield, let us now construct a $q$-covariant deri\-varive, $D$,
that is, a differential operator that obeys
\begin{eqnarray}
&[ D \, , \,Q ]_q = 0,&  \label{qcomd} \\
&D\,(\Delta X) = \Delta \, (D X).&  \label{ddeltax}
\end{eqnarray}
We could try for $D$ the same structure that appears in the $q$-SUSY
generator, i.e., to take $D = q^{\alpha} \theta^2 \partial_t + a q^{\beta}
\partial_{\theta}$, ($\alpha,\beta = 1,2,3; a \in C$). However,
it turns out not to be possible to find an operator with this structure and
quommuting with $Q$. The only operator that obeys (\ref{ddeltax}) is $Q$
itself, but it obviously does not obey (\ref{qcomd}).

To construct the coordinates of level $3$ permit us to introduce two
differential operators, $%
\partial_{\theta}$ and $\delta{_\theta}$.  Using the second one it is
possible to show that the operator
\begin{equation}
D = \theta^2 \partial_t + q \delta_{\theta},
\end{equation}
satisfies the conditions (\ref{qcomd}) and (\ref{ddeltax}).

As in the supersymmetric case, the component fields can be defined by
projecting the superfield on different sectors, using the covariant
derivatives on $\theta = 0$.
\begin{eqnarray}
X |_{\theta=0} &=& x ,  \nonumber \\
D X |_{\theta=0} &=& \psi^{(2)} ,  \nonumber \\
D^{2} X |_{\theta=0} &=& - \psi^{(1)} .
\end{eqnarray}
 From now on, we will neglect the subscript $\theta = 0$.

We also notice some relations between different powers of $D$ and $Q$, that
will become useful later
\begin{eqnarray}
D \; . \; | &=& q^2 \, Q \; . \; |,  \nonumber \\
D^2 \; . \; | &=& q \, Q^2 \; . \; |,  \nonumber \\
D^3 \; . \; | &=& - \, \partial_t \; . \; |.
\end{eqnarray}

Besides the above-defined bosonic superfield, we can also construct  sectors
one and two superfields. Their $\theta$ expansion can be taken to be
\begin{equation}
\Lambda^{(1)} = \lambda^{(1)} + \theta A + q \theta^2 \lambda^{(2)},
\end{equation}
and
\begin{equation}
\Xi^{(2)}(t) = \xi^{(2)} + q \theta \xi^{(1)} + {\theta}^2 F \; ,
\end{equation}
where the superscripts indicate the sectors to which the fields belongs, and
$A$ and $F$ are bosonic fields.

The dimension of the $q$-superfield $\Xi^{(2)}$ is taken to be $\frac{2}{3}$%
, its bosonic component $F$ being dimensionless and, as we will see later,
behaving as an auxiliary field. We cannot, however, take the dimension of
the $q$-superfield $\Lambda^{(1)}$ to be $\frac{1}{3}$, since this would
imply a negative dimension for the component field $\lambda^{(2)}$. Thustake
its dimension to be $\frac{4}{3}$. This, however, will produce
different equations of motion for its quermionic components, as we will see
in the next section.

We assume that the fields $\xi^{(j)}$ have the same behaviour as $\psi^{(j)}$
with respect the quommutations relations with each other, with $\theta $ and
with $\varepsilon$.



\section{Examples of Superactions}

In this section, we are going to make a general discussion about simply
quadratic actions that are functions of the $q$-superfields introduced in
the previous  section and give some examples of them.

The action for a generic superfield $\Phi$ must be of the form
\begin{equation}
S = \int dt d\theta \; {\cal P} (\Phi, \dot\Phi, D\Phi,D^2\Phi),
\label{acaogeral}
\end{equation}
where ${\cal P}$ is a polynomial in $\Phi$ and its derivatives. ${\cal P}$
must behave like $\theta^{2}$, belonging to the sector two (since $\int
d\theta = {\partial_\theta}^{2}$, and $S$ is scalar), and since the measure
has mass dimension $\frac{-1}{3}$  and $S$ is dimensionless, its dimension
must be $\frac{1}{3}$.

By comparing the expression for the covariant derivative and the $\theta$%
-integration, we notice the rule
\begin{equation}
\int d\theta = q^2 D^2 |.  \label{dtetaeD}
\end{equation}

Let us now perform a transformation on the action
\begin{equation}
\Delta S = \int dt d\theta \, \Delta {\cal P}(\Phi, \dot\Phi, D\Phi,
D^2\Phi),
\end{equation}
since the Jacobian is one, which can be seen by the $t$-independence of  the
$\theta$ translation. Since ${\cal {P}}$ is a superfield, its variation is
of the form of eq.(\ref{deltafi}). Using this and eq.(\ref{dtetaeD}), we
arrive  at the conclusio\begin{equation}
\Delta S = - \, q \, \, \varepsilon \; \int dt \, \partial_t {\cal P}.
\label{vardeS}
\end{equation}
and the transformations eq.(\ref{vardetetaet}) generates symmetries of the
action.

Let us now write an action of the q-superfields $X$, $\Lambda^{(1)}$ and $%
\Xi^{(2)}$ defined in Section 3, and compute their equations of motion. We
begin  with the bosonic superfield $X$. Its quadratic action is
\begin{equation}
S_X = - \frac{m}{2} \, \int dt d\theta \; q^2 \; (D^2X)(D^2X),
\label{acaopl}
\end{equation}
where $m$ is a commuting mass parameter. By explicit computation of its $%
\theta$ integral, or by use of eq.(\ref{dtetaeD}), this action can be writen
down in in components as
\begin{equation}
S_X = m \int dt \left( \frac{1}{2} {\dot{x}^2} - 2 q \dot{\psi}^{(2)}
\psi^{(1)} \right),
\end{equation}
where the difference with the Section 1 action is due to the different
initial superactions in these cases.

Although the variational calculus of the quermionic coordinates presents
several difficulties to overcome (for instance, how to do the  variation
with respect to a quermion), it is clear that the equation of motion arising
from the above Lagrangian is, up to multiplicative factors $D \dot{X} = 0 $,
giving in components $\ddot{x} = {\dot{\psi}^{(j)}} = 0 $ ($j=1,2$).
Computing its $q$-supersym\-metric variation, we obtain
\begin{equation}
\Delta S_X = q \, \epsilon \int dt \frac{\partial {\psi^{(1)}}^2}{\partial t} ,
\label{deltaacaopl}
\end{equation}

We notice that the action given by eq.(\ref{acaopl}), its variation eq.(\ref
{deltaacaopl}) and the variation of the component fields eq.(\ref{deltacomp}%
) are, up to factors, equal to eqs.(\ref{act}), (\ref{deltaact}) and (\ref
{deltacomp1}) respectively, recalling that the presence of such cocicle-type
factors was because the quommutation homogeneity assumption had been used.
Thus we see that  the $q$-superfield $X$ describes the dynamics of a free
particle partners.

The quadratic action for the $q$-superfield $\Lambda^{(1)}$ is
\begin{equation}
S_{\Lambda} = - \frac{m}{2} \int dt d\theta \, \, (\dot{\Lambda}^{(1)})^2.
\end{equation}
By convenience the mass parameter was taken to be the same as in the $X$
action. In components, the action turns out to be
\begin{equation}
S_{\Lambda} = \frac{m}{2} \int dt \, (\dot{A}^2 + 2q \dot{\lambda}^{(2)}
\dot{\lambda}^{(1)}).
\end{equation}
It is interesting to notice that the equation of motion for $\Lambda^{(1)}$,
obtained from its action, $\ddot{\Lambda}^{(1)} = 0, $ gives in component $%
\ddot{A} = \ddot{\lambda}^{(i)} = 0$. Thus this $q$-superfield also
represents a free particle, but its quermionic  partners obey an equation of
motion that is of second order in the time  derivative, whereas in the case
of $q$-superfield $X$ it is of first order. The $q$-supersymmetric variation
of the $S_{\Lambda}$ is
\begin{equation}
\Delta S_{\Lambda} = \varepsilon \, \int dt \frac{\partial (\dot{\Lambda}%
^{(1)})^2}{\partial t} .
\end{equation}

We now consider the quadratic action for the q-superfield $\Xi^{(2)}$. It is
\begin{equation}
S_{\Xi} = m \int dt d\theta \; \, (D \Xi^{(2)})^2 .  \label{acaocsi}
\end{equation}
In component fields, the action reads
\begin{equation}
S_{\Xi} = m \int dt \, [ 2 q \,\dot{\xi^{(2)}} \xi^{(1)} + F^2].
\end{equation}
The equation of motion for $\Xi^{(2)}$ is $D^2 \Xi^{(2)} = 0 , $ giving $F =
\dot{\xi}^{(j)} = 0$, meaning, as it was anticipated, that  the bosonic
coordinate $F$ is an auxiliary one. The variation of $S_\Xi$ is
\begin{equation}
\Delta S_\Xi = - \epsilon \int dt \, \frac{\partial {\xi^{(1)}}^2}{\partial t%
}.
\end{equation}

The superfields $X$ and $\Xi^{(2)}$ can have a quadratic action with a mixed
term
\begin{equation}
S_{X \Xi} = m \omega \int dt d\theta q^2 \; X \Xi^{(2)},  \l\end{equation}
where $\omega$ has a $\mbox{mass}^{-1}$ dimension. In components we write
this action as
\begin{equation}
S_{X \Xi} = m \omega \int dt \left( \, F x + q^2 \psi^{(1)} \xi^{(2)} + q
\psi^{(2)} \xi^{(1)} \right). \label{acaomix}
\end{equation}
Summing up the actions (\ref{acaopl}), (\ref{acaocsi}) and  (\ref{acaomix}) $%
S_{HO} = S_X + S_{\Xi} + S_{X \Xi}, $ and its bosonic part is
\begin{equation}
S_{HO} = \int dt \, m (\frac{1}{2} {\dot{x}}^{2} + \frac{1}{2} F^2 + \omega
F x).
\end{equation}
Computing the equation of motion of the auxiliary field $F$ and
reintroducing it in the action, it becomes
\begin{equation}
S_x = \int dt \left[ \frac{1}{2} m {\dot{x}}^2 - \frac{1}{2} m \omega x
\right],
\end{equation}
which is the action for the harmonic oscillator.
%
%

\section{Conclusions}

In this letter, we presented a generalization of some supersymmetric
classical mechanical models where the superspace has a non-commu\-ting
coordinate nilpotent of order $3$, and the commutation relations among the
several objects of the model are deformed by powers  of a parameter $q$.
Translations on the $q$-superspace induce transformations on the fields, and
the operatorial realization of the supersymmetric generator is obtained by a
suitable Taylor expansion. The covariant derivative was also introduced,  in
which we used a second kind of partial paragrassmannian derivative. Spite
the supersymmetric structure similarity, we are facing a slightly different
situation. In fact because of the presence of two derivatives, like the
forward and the backward one, it resembles a lattice approach. In a very
recent and nice work \cite{Mac2}, the authors showed the r\^oles playing by
the covariant derivative $D$ and the symmetry generator $Q$, present in this
work, are the left and right action of $G_3$ group.

After introducing superfields belonging to different sectors, we were able to
construct quadratic actions for each one. These actions are, up  to total
derivatives, invariant under the $q$-supersymmetric transformations.  Using
a na\"\i ve approach, it is possible to extract from these  actions the
equations of motion since there is no, up to now, a  well-defined
differential calculus on these quermionic coordinates. We intend to  discuss
this  subject in a forthcoming publication. We also showed that imposing the
``on-shell'' constraint to the auxiliary fields, it is possible to  get the
harmonic oscillator as a bosonic sector of a simple suitable  linear
combination of the actions.

It should also be interesting to study this formulation from a field
theoretical point of view, in particular in the $(2+1)$-dimensional case. We
might also try to understand if such fields are representations of some $q$%
-deformed algebra, either a $q$-Poincar\'e or a $q$-Clifford one.

\vspace{1cm} Acknowledgements: The authors would like to thank J.A.
Helay\"el-Neto and Marco A. R. Monteiro for discussions. The authors also
thanks to CNPq  and FAPERJ for financial support and for the DCP/\-CBPF
hospitality.

\end{document}